\documentclass[preprint1,12pt]{aastex}

\usepackage{latexsym,amssymb}
\usepackage{amsmath}
\usepackage{natbib}

\usepackage[colorlinks=true,linkcolor=blue,citecolor=blue]{hyperref}

\slugcomment{Draft \today}
\shorttitle{sGRB Afterglow Diversity}
\shortauthors{Holcomb et al.}

\begin{document}

\title{Diversity of Short Gamma-Ray Burst Afterglows from Compact Binary Mergers Hosting Pulsars}
\author{Cole Holcomb\altaffilmark{1,2}, Enrico Ramirez-Ruiz\altaffilmark{1}, Fabio De Colle\altaffilmark{1,3},  and Gabriela Montes\altaffilmark{1}}
\altaffiltext{1}{Department of Astronomy and
  Astrophysics, University of California, Santa Cruz, CA
  95064}
  \altaffiltext{2}{Department of Astrophysical Sciences, Princeton University, 4 Ivy Lane, Princeton, NJ 08540, USA}
\altaffiltext{3}{Instituto de Ciencias Nucleares, Universidad Nacional Aut{\'o}noma de M{\'e}xico, A. P. 70-543 04510 D. F. Mexico}

\begin{abstract}
Short gamma-ray bursts (sGRBs) are widely believed to result from the mergers of compact binaries. This model predicts an afterglow that bears the characteristic signatures of a constant, low density medium, including a smooth prompt-afterglow transition, and a simple temporal  evolution. However, these expectations are in conflict with observations for a non-negligible fraction of sGRB afterglows. In particular, the onset of the afterglow phase for some of these events appears to be delayed and, in addition, a few of them exhibit late-time rapid fading in their lightcurves. 
We show that these peculiar observations can be explained independently of ongoing central engine activity if some sGRB progenitors are compact binaries hosting at least one pulsar. 
The Poynting flux emanating from the pulsar companion can excavate a bow-shock cavity surrounding the binary. If this cavity is larger than the shock deceleration length scale in the undisturbed interstellar medium, then the onset of the afterglow will be delayed.
Should the deceleration occur entirely within the swept-up thin shell, a rapid fade in the lightcurve will ensue.  We identify two types of pulsar that can achieve the conditions necessary for altering the afterglow: low field, long lived pulsars, and high field pulsars. We find that a sizable fraction ($\approx $20-50\%) of low field pulsars are likely to reside in neutron star binaries based on observations, while their high field counterparts are not. Hydrodynamical calculations motivated by this model are shown to be in good agreement with observations of sGRB afterglow lightcurves.
\end{abstract}

  
\section{Introduction} 

The leading progenitor models for short duration gamma-ray bursts (sGRBs)
are merging compact binaries \citep[for a review, see][]{2007NJPh....9...17L} consisting of either two neutron stars (NSs) \citep{1986ApJ...308L..43P,1989Natur.340..126E,1992ApJ...395L..83N} or a black hole (BH) with a NS companion \citep{1976ApJ...210..549L}.
 Observations of sGRB host galaxies are broadly consistent with the expected host morphologies, offsets, and delay time distributions of compact mergers \citep{2006ApJ...642..989P,2010ApJ...708....9F,2013ApJ...769...56F,2010ApJ...722.1946B,2014arXiv1401.7986B},
 and the recent discovery of an r--process powered kilonova
 \citep{1998ApJ...507L..59L,2010MNRAS.406.2650M,2011ApJ...736L..21R,2013ApJ...773...78B,2013ApJ...775...18B,2013ApJ...775..113T,2014MNRAS.439..757G} associated with GRB130603B has given further credence to  this model \citep{2013Natur.500..547T,2013ApJ...774L..23B}.

Compact binaries are expected to receive large ``natal kicks" due to asymmetries manifested by the mechanisms of NS/BH formation, resulting in mean velocities of $\approx 180$ km s$^{-1}$ for NS/NS binaries and $\approx 120$ km s$^{-1}$ for NS/BH binaries \citep{1998ApJ...499..520F,2010ApJ...721.1689W,2014arXiv1401.7986B}.
Kicks of these magnitudes imply that compact binaries quickly leave behind their formation site, and often reach large distances from their host galaxies \citep{2010ApJ...725L..91K}.  It was therefore anticipated that compact binary mergers would commonly occur in constant low density media characteristic of interstellar or intergalactic space \citep{2002ApJ...570..252P}.
The interaction between an ultra-relativistic blastwave and its surrounding medium was predicted to produce a dim and smooth afterglow in a broad range of wavelengths shortly following the prompt emission \citep{1992MNRAS.258P..41R,2001ApJ...561L.171P}. Observations have shown, however, that the onset of the afterglow emission  can sometimes be postponed from tens to hundreds of seconds following the prompt emission. In addition, some events have exhibited an abrupt turn-off to a rapid fading in the afterglow lightcurve at late times \citep{2013MNRAS.430.1061R}. These facts have currently been interpreted as arising from late time energy injection via the formation of a long lived central engine at the merger site \citep{2008MNRAS.385.1455M,2013MNRAS.430.1061R,2013MNRAS.431.1745G}. 

In this \emph{Letter} we offer an alternative explanation. As noted above, it has long been known that the afterglow properties   depend  heavily on the conditions of the circumburst medium. If a  sGRB progenitor hosts a  pulsar, then a Poynting flux dominated cavity will form in its vicinity. For a compact binary system with a center of mass velocity exceeding a few tens of kilometers per second, this evacuated structure  will manifest itself as a bow-shock cavity in which a massive thin shell separates the circumbinary medium from the interstellar medium. If this bubble grows to sufficiently large size, then the deceleration of the sGRB jet, and thus the onset of the afterglow, will be significantly  delayed. Additionally, should the shock deceleration occur predominantly within the bow-shock  thin shell, the afterglow is predicted to quickly  fade. In this way, we can account for some of the diversity of observed afterglow behaviors without invoking the existence of a long lived central engine. In Section \ref{sec:cav} we describe the conditions necessary for the production  of large bow-shock cavities  in systems hosting pulsars. In Section \ref{sec:aft} we derive the observational implications for sGRBs, present relativistic hydrodynamical simulations of the sGRB jet propagation taking place within such pulsar bow-shock cavities, and compare with observed data. Finally, in Section \ref{sec:dis} we summarize our results and discuss their implications for sGRBs arising from neutron star mergers hosting pulsars.

\section{Properties of Pre-Explosive Cavities}\label{sec:cav}

\begin{figure}[]
\centering\includegraphics[width=\linewidth,clip=true]{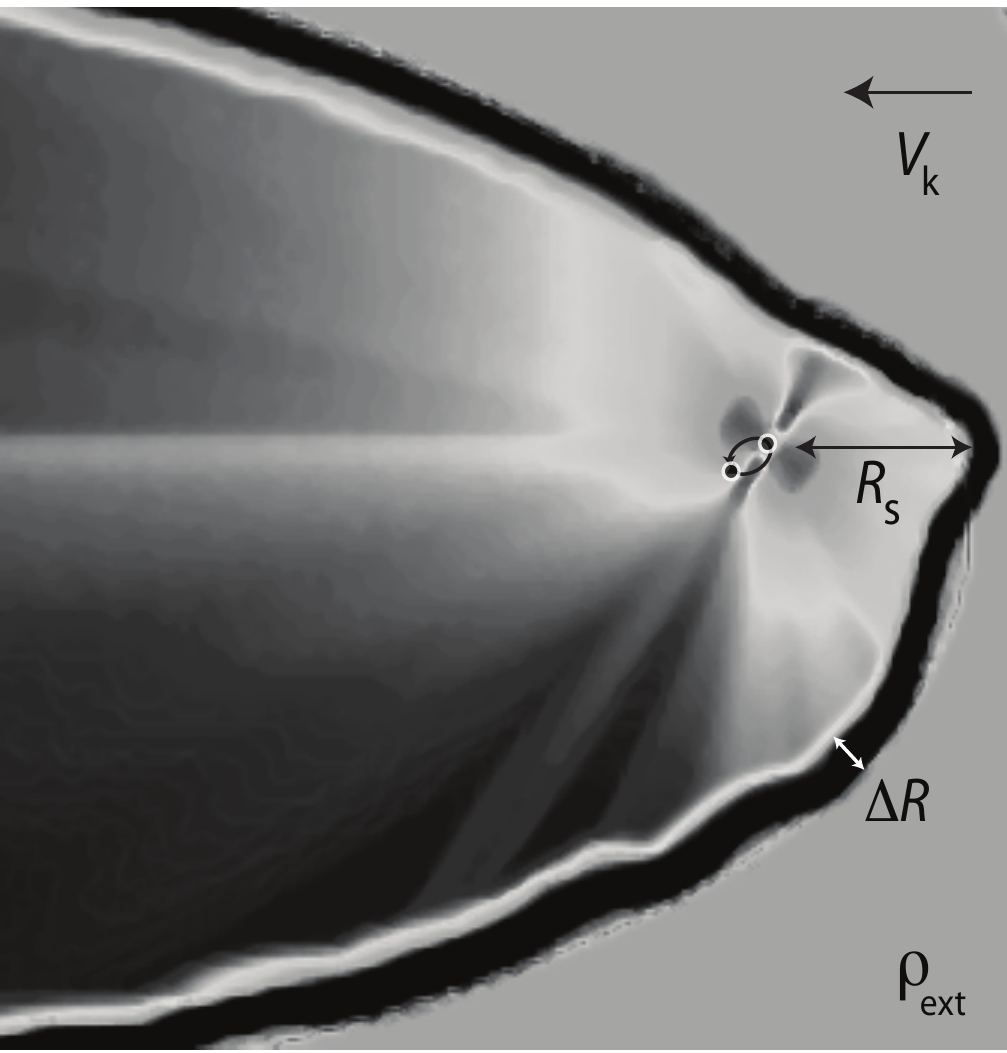}
\caption{The bow-shock parameters; adapted from \cite{2007MNRAS.374..793V}. The figure depicts the environment expected for sGRBs occuring inside a cavity of size $R_{\rm s}$  inflated by a pulsar in the  binary. The binary  travels through a medium of density $\rho_{\rm ext}$ with a velocity $V_{\rm k}$.  $R_{\rm s}$ designates  the distance at which  the external ram pressure balances the pulsar wind ram pressure.}
\label{fig:diag}
\end{figure}

A binary with a  pulsar companion with isotropic luminosity $L$, traveling through some medium of density $\rho_{\rm ext}$ with velocity $V_{\rm k}$ will form a bow-shock with characteristic size $R_{\rm s}$ given by \citep{1996ApJ...459L..31W,2000ApJ...532..400W}
\begin{eqnarray}
R_{\rm s}  &=&  \left({L\over 4\pi c \rho_{\rm ext} V_{\rm k}^2}\right)^{1/2} \nonumber \\
  &\approx&  4.0 \times 10^{16}\; L_{35}^{1/2}n_{\rm ext}^{-1/2} V_{{\rm k},2}^{-1} \; {\rm cm},
\label{eq:rbow}
\end{eqnarray}
where $L_{35} = L/(10^{35} \text{ erg s$^{-1}$})$, $V_{{\rm k},2} = V_{\rm k}/(10^2 \text{ km s}^{-1})$, and $n_{\rm ext}$ is the external particle density. This length scale, called the standoff length, has been shown to be in reasonable agreement with MHD calculations of the shock structure even when the luminosity is anisotropic and nonaxisymmetric \citep[see Fig. \ref{fig:diag};][]{2007MNRAS.374..793V}. Equation \ref{eq:rbow} becomes exact in the limit of efficient cooling, but realistic systems form a thin shell of width $\Delta R$ between the bow-shock and an inner contact discontinuity (Fig.\  \ref{fig:diag}).

For a merging binary with at least one pulsar, there are two channels of luminosity production that can carve a sizable bow-shock cavity. The first channel is from the traditional magnetic dipole model of a pulsar,
\begin{eqnarray}
L_{\rm p}  &=& \frac{6c^3I^2}{B^2 R^6 \tau_{\rm p}^2} \nonumber \\
		&\approx& 1.4\times 10^{35}\; B_{9}^{-2} \tau_{{\rm p}, 9}^{-2} \text{erg s$^{-1}$},
\label{eq:lp}
\end{eqnarray}
where $\tau_{\rm p} = P/\dot P$ is the pulsar spin-down time scale ($\tau_{{\rm p},9} = \tau_{\rm p}/10^9 \text{ yr}$), $P$ is the pulsar period, $B$ is the pulsar magnetic field strength ($B_{9} = B/10^{9} \text{ G}$), $R$ is the pulsar radius, and $I = 2MR^2/5$ is the moment of inertia. In equation \ref{eq:lp}, and throughout this work, we have taken $R=1.2\times 10^6$ cm and $M = 1.4M_{\odot}$ as fiducial NS parameters.

The second channel is the energy dissipation due to the torque on the binary system by the pulsar magnetic field \citep{2012ApJ...757L...3L,2012ApJ...755...80P},
\begin{align}
L_{\rm B} &\approx 7.4\times 10^{36}\; \zeta_\phi B_{9}^2 a_{30}^{-13/2}\; \text{ erg s$^{-1}$},
\end{align}
where $\zeta_\phi$ is the azimuthal twist (hereafter taken to be equal to the $\zeta_\phi \approx 1$ upper bound). The orbital separation $a$ ($a_{30} = a/30 \text{ km}$), whose evolution is driven by gravitational wave emission, is given by
\begin{align}
a_{30} &\approx \left(\frac{\tau_{\rm GW}}{1.2 \times 10^{-2}\; \text{s}}\right)^{1/4},
\end{align}
where $\tau_{\rm GW}=a/\dot a$ is the merging time scale \citep{2012ApJ...757L...3L}.

To calculate the size of the cavity produced by the system, we substitute $L_{\rm tot} = L_{\rm p} + L_{\rm B}$ into equation \ref{eq:rbow}. However, because the energy emitted propagates to the bow-shock interface in finite time, there is a lag of $R_{\rm s}/c$  before $R_{\rm s}$ can respond to changes in the luminosity at the source. We therefore make the transformation $\tau_{\rm GW} \rightarrow \tau_{\rm GW} + R_{\rm s}/c$ for our calculations.

In the following we will compare $R_s$ to the length scale of sGRB shock deceleration in the standard fireball model \citep{1992MNRAS.258P..41R},
\begin{align}
R_{\rm dec} & = \bigg ( \frac{3E}{4\pi \Gamma_0^2 c^2 m_p n_{\rm ext}} \bigg )^{1/3} \nonumber \\
& \approx  5.6 \times 10^{15}\; E_{50}^{1/3}n_{\rm ext}^{-1/3}\Gamma_{300}^{-2/3}  \; {\rm cm},
\label{eq:rdec}
\end{align}
where $E_{50} = E/(10^{50} \text{ erg})$ is the  isotropic-equivalent sGRB energy output and $\Gamma_{300} = \Gamma_0/300$ is the bulk Lorentz factor. As we will see in Section \ref{sec:aft}, the behavior of the sGRB afterglow is modified if $R_{\rm s} \gtrsim R_{\rm dec}$ since, in the presence of a cavity, the swept-up mass required to cause the deceleration of a sGRB jet is not encountered until after $R_{\rm s}$.

\begin{figure*}[]
\centering\includegraphics[width=\linewidth,clip=true]{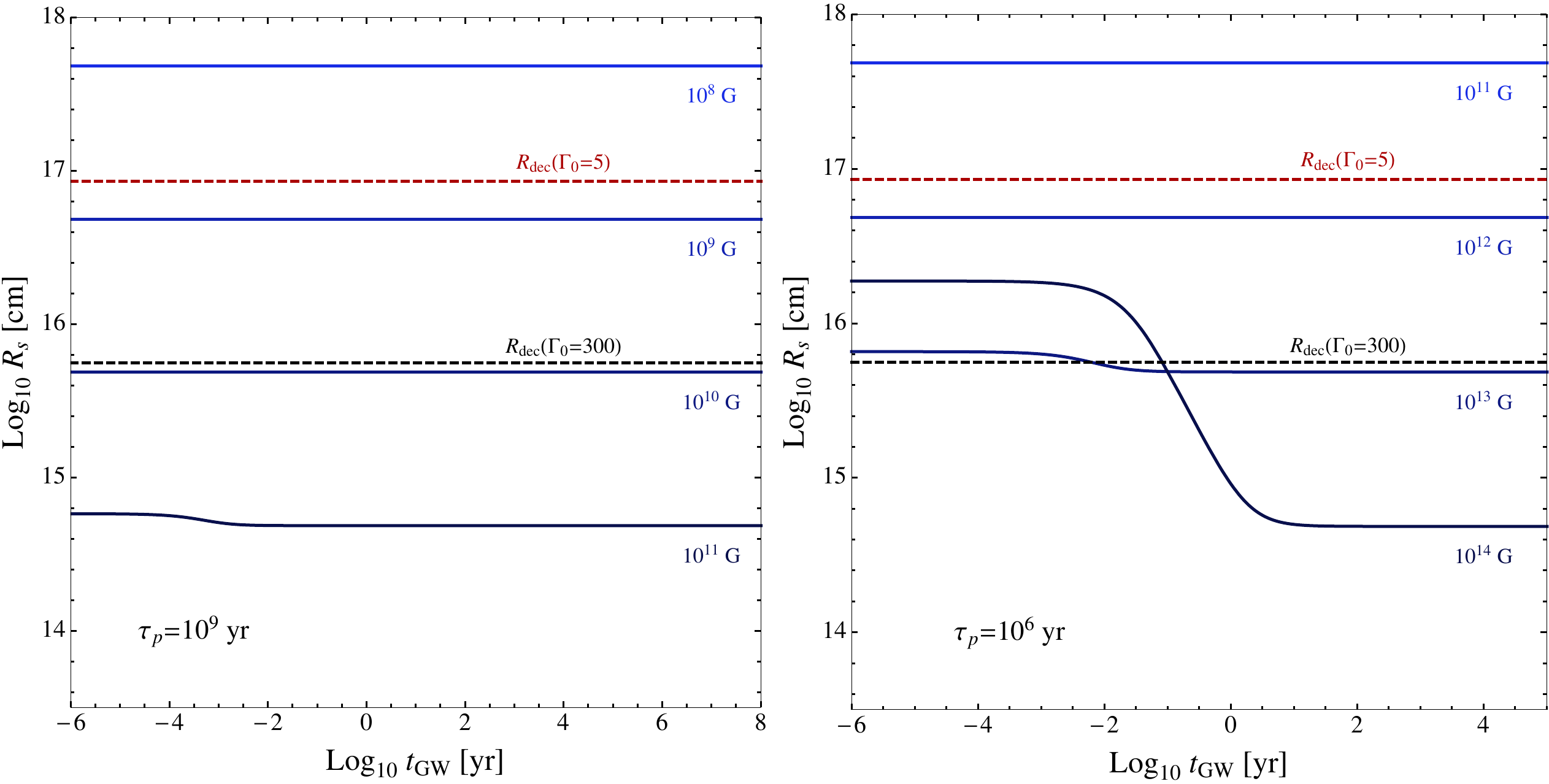}
\caption{The growth of the pre-explosive cavities. For a fixed pulsar spin-down time $\tau_{\rm p}=10^9$ yr (left) and $10^7$ yr (right), we show the evolution of  $R_{\rm s}$ as a function of the time to merger $\tau_{\rm GW}$ for a range of pulsar magnetic fields. These calculations assume $V_{\rm k}=100$ km s$^{-1}$. We have marked the radii for shock deceleration in both the highly relativistic ($\Gamma_0=300$; {\it black} dashed line) and mildly relativistic ($\Gamma_0=5$; {\it red} dashed line) limits, where $E_{50} = 1$ and $n_{\rm ext} = 1\,{\rm cm^{-3}}$. On time scales  $\lesssim \tau_{\rm p}$, the spin-down luminosity $L_{\rm p}$ remains roughly constant and is the dominant contribution to $R_{\rm s}$. The magnetic torque only becomes a significant luminosity source in the last few weeks prior to merger, and only for high magnetic fields.}
\label{fig:growth}
\end{figure*}

In Fig.\ \ref{fig:growth} we show the evolution of $R_{\rm s}$ as a function of the time to merger $\tau_{\rm GW}$ for a fixed $\tau_{\rm p} \gtrsim \tau_{\rm GW}$. $L_{\rm p}$ is roughly constant over time scales $\lesssim \tau_{\rm p}$, so $R_{\rm s}$ remains unchanged unless the pulsar magnetic field is sufficiently high to drive a significant increase in $L_{\rm B}$ (equation \ref{eq:lp}) at small orbital separations. We set $\tau_{\rm p} = 10^7$ yr and $10^9$ yr, roughly corresponding to the range of  expected sGRB delay times \citep{2006A&A...453..823G,2010ApJ...725.1202L,2013arXiv1311.2603B,2014arXiv1401.7986B}.
Within this range,  magnetic fields $\gtrsim  10^{13}$ G are required for $L_{\rm B}$ to be the dominant contribution to $R_{\rm s}$.  In all other cases, $L_{\rm p}$ determines the size scale of the bow-shock cavity. For comparison, we plot $R_{\rm dec}(\Gamma_0=5)$ ({\it red} dashed line) and $R_{\rm dec}(\Gamma_0=300)$ ({\it black} dashed line) in Fig.\ \ref{fig:growth}. We note that $R_{\rm s} > R_{\rm dec}$ for a large range of  realistic pulsar parameters; the reader is referred to Section~\ref{sec:dis} for a detailed account of the fraction of neutron star binaries expected to satisfy this relation. 

\begin{figure}[]
\centering\includegraphics[width=\linewidth,clip=false]{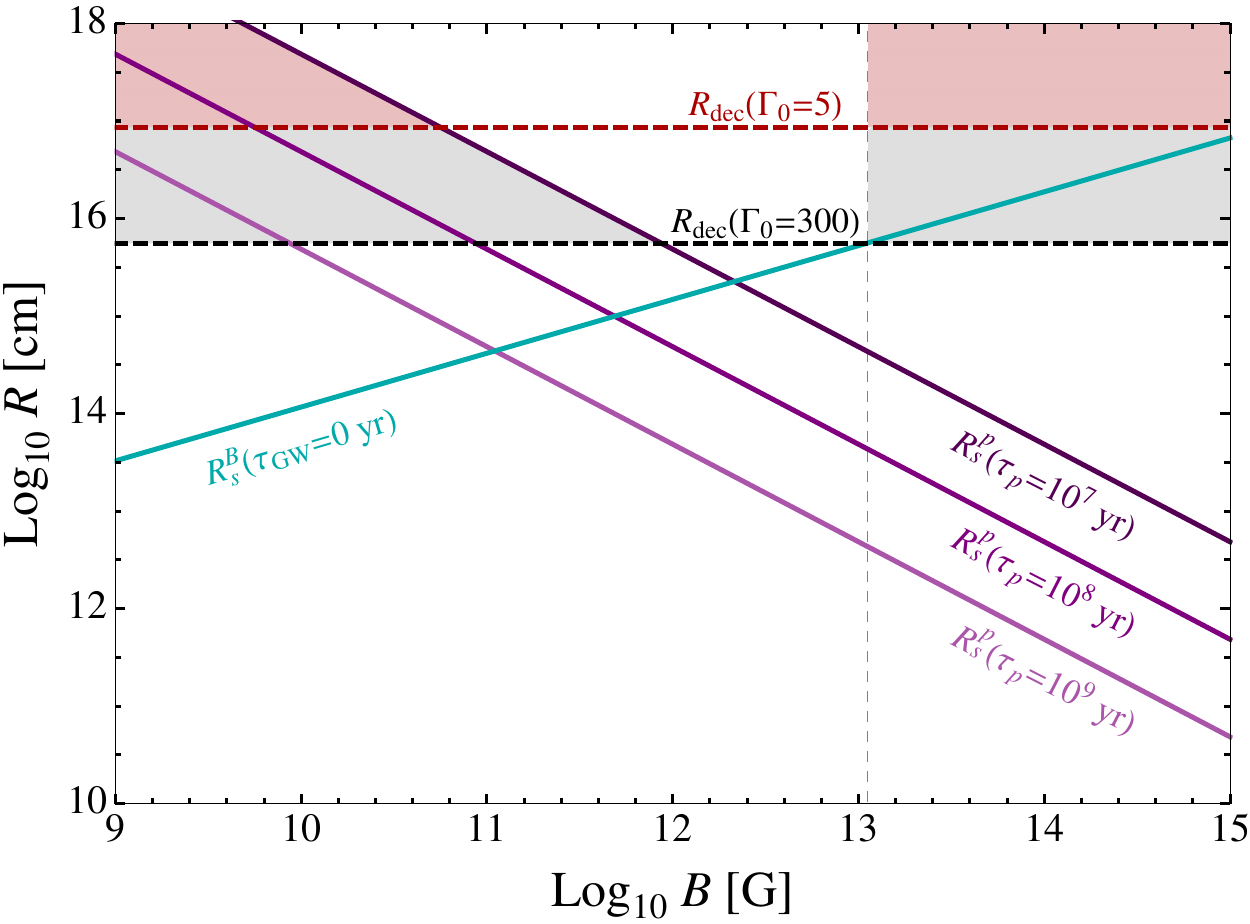}
\caption{The individual contributions to the production of the pre-explosive cavities. The contributions to the standoff radius $R_{\rm s}\approx\max(R_{\rm s}^{\rm p}$,\,$R_{\rm s}^{\rm B}$) are plotted separately against the pulsar magnetic field. The {\it purple} lines indicate the contributions from the pulsar spin-down luminosity for given lifetimes $\tau_{\rm p}$, and the {\it cyan} line denotes the contribution from the magnetic torque on the binary at the time of the merger $\tau_{\rm GW} = 0$. We assume  $V_{\rm k}=100$ km s$^{-1}$.  As in Fig.\ \ref{fig:growth}, we show $R_{\rm dec}(\Gamma_0=300)$ ({\it black} dashed line) and $R_{\rm dec}(\Gamma_0=5)$ ({\it red} dashed line). The {\it grey} and {\it red} shaded regions delimit the parameters of interest; long lived, weakly magnetized pulsars and strongly magnetized pulsars regardless of spin-down lifetime.}
\label{fig:size}
\end{figure}

We can compare the relative importance of each luminosity contribution to $R_{\rm s}$ by separating it into $R_{\rm s}^{\rm B}= R_{\rm s}(L_{\rm B})$ and $R_{\rm s}^{\rm p}=R_{\rm s}(L_{\rm p})$. In Fig.\ \ref{fig:size} we plot these radii as a function of the pulsar magnetic field $B$ for $\tau_{\rm GW} = 0$ and fiducial values of $\tau_{\rm p}$. The dominant contribution to $R_{\rm s}$ is from $L_{\rm p}$ for nearly all realistic values of $B$ and $\tau_{\rm p}$, while $L_{\rm B}$ dominates for $B>10^{13}$ G, irrespective of $\tau_{\rm p}$ (assuming no field decay; see Section~\ref{sec:dis}).

\section{sGRB Afterglows in Pre-Explosive Cavities}\label{sec:aft}
In the standard fireball model for GRB afterglows, an effective energy release occurs at observer time \citep{1992MNRAS.258P..41R,2006ApJ...642..354Z}
\begin{align}
t_{\rm dec} &=\bigg({1+z\over 2}\bigg ){R_{\rm dec}\over \Gamma_0^2c}   \nonumber \\
& \approx  2.1 \; {\bigg({1+z\over 2}\bigg)} E_{50}^{1/3}n_{\rm ext}^{-1/3} \Gamma_{300}^{-8/3} \; {\rm s}
\label{eq:tdec}
\end{align} 
following the prompt emission. This is the time it takes for the outgoing sGRB shockwave to sweep up a mass $M = E/\Gamma^2_0 c^2\approx 6.2\times 10^{-10} E_{50}\Gamma_{300}^{-2}M_\odot$.  If, however, $R_{\rm s}\gtrsim R_{\rm dec}$, then the mass required to decelerate the ejecta is pushed out to $R_{\rm s}$ by the Poynting flux emanating from the  pulsar companion. In this case, the onset of the afterglow will be delayed to
\begin{align}
t_{\rm dec, cav}&= \left({R_{\rm s}\over R_{\rm dec}}\right) t_{\rm dec} \nonumber \\
 &\approx  15.0 \; {\bigg ({1+z\over 2}\bigg )}  L_{35}^{1/2}n_{\rm ext}^{-1/2} V_{{\rm k},2}^{-1}\Gamma_{300}^{-2} \; {\rm s}
 \label{eq:tcav}
\end{align} 
in the observer frame.
The fulfillment  of  the condition $R_{\rm s}\gtrsim R_{\rm dec}$ would naturally give rise to a wide range of afterglow behavior. Since $R_{\rm s}$ can in some cases greatly exceed $R_{\rm dec}$ (Fig.~\ref{fig:size}), afterglows can be expected to trail the prompt emission by tens to hundreds of seconds, depending on the properties of the pulsar companion. These delays are consistent with observations \citep{2013MNRAS.430.1061R,2013MNRAS.431.1745G}.

To illustrate the afterglow  properties of sGRBs originating inside pulsar cavities we make use of the {\it Mezcal} special relativistic adaptive mesh refinement hydrodynamics code, a parallel, shock capturing code routinely used to simulate the propagation of relativistic jets \citep{2012ApJ...760..103D,2012ApJ...751...57D}.
A detailed description of the numerical framework, afterglow lightcurve calculation  and  tests of the SRHD implementation are presented in \citet{2012ApJ...746..122D}. 
For this discussion we shall assume the blastwave is adiabatic
and effectively spherical.

Usually in the study of sGRB afterglows from compact binary mergers one considers expansion into a uniform medium. However, since the wind of the pulsar meets the interstellar medium  at some point, the density structure is more complex. In the first case depicted in Fig.\ \ref{fig:aft} ({\it red} curve), the  pre-explosive pulsar wind  is weak enough that the ejecta is not significantly slowed down by the time it expands beyond the pulsar wind cavity $R_{\rm s} \ll R_{\rm dec}(\rho_{\rm ext})$, so that we expect the blast wave evolution to take place in a constant density medium. This phase ends when the energy is  shared with so much material that the blast wave becomes sub-relativistic. Beyond this point,  the blastwave  slowly transitions  into the classical Sedov-Taylor  evolution, leading to a steeper decline in the lightcurve \citep{1998ApJ...497..288W,2005ApJ...631..435R}.

In the presence of a  more luminous  pulsar companion, the shock front will expand unimpeded within the pulsar wind cavity until it reaches the density discontinuity placed at $R_{\rm s}=2.5R_{\rm dec}(\rho_{\rm ext})$ ({\it light blue} curve of Fig.\ \ref{fig:aft}).  Over the typical time of observation of a sGRB, the impact of the blastwave with the thin shell will produce a prominent  extended feature in the observed  afterglow emission. The thin shell  in this example is not massive  enough to slow down the ejecta to non-relativistic speeds before the shock reaches the interstellar medium. In the last  case depicted in Fig.~\ref{fig:aft} ({\it dark blue} curve),  the cavity {\it is} large enough to slow down the ejecta to non-relativistic speeds before reaching the interstellar medium, as would occur for systems with large luminosities or small kick velocities. In this case, we expect the blastwave evolution as we see it to take place entirely within the thin shell, which has been placed at $R_{\rm s}=5R_{\rm dec}(\rho_{\rm ext})$. If this occurs, then it is expected that the afterglow will exhibit a rapidly declining lightcurve (like in  the X-ray afterglows  of GRB 051210 and GRB 060801; see Fig. \ \ref{fig:aft}), which can be even more pronounced in two dimensional calculations when  the Lorentz factor of the flow drops below the inverse of the jet opening angle \citep{2013ApJ...773....2G}. A steeper decline  \citep[like in the X-ray afterglow of  GRB 090515;][]{2013MNRAS.430.1061R}  might  also be expected if the magnetic field energy density, here assumed to be a fixed fraction $\epsilon_B$ of the downstream energy density of the shocked fluid, drastically changes as the blastwave expands beyond the thin-shell (otherwise such steep declines might be difficult to reconcile with the simple model outlined here).  This study reveals how the  properties of sGRB afterglows from neutron star mergers depend sensitively  on whether or not one of the binary members is a pulsar.

\section{Discussion}\label{sec:dis}
The key to delaying the onset of a sGRB afterglow lies in the relation $R_{\rm s}\gtrsim R_{\rm dec}$, and therefore it is natural to examine the range of conditions  for which this constraint is satisfied. Taking the ratio of equations \ref{eq:rbow} and \ref{eq:rdec}, we find
\begin{align}
\left(\frac{R_{\rm s}}{R_{\rm dec}}\right) \approx 7.1\, L_{35}^{1/2} \Gamma_{300}^{2/3} E_{50}^{-1/3} n_{\rm ext}^{-1/6} V_{{\rm k},2}^{-1}.
\label{eq:ratio}
\end{align}
This ratio is modestly dependent on the properties of the pulsar companion  and the bulk Lorentz factor but highly insensitive to $n_{\rm ext}$ and $E$. 
It is clear from equation \ref{eq:ratio} that a lower kick velocity can ensure $R_{\rm s}>R_{\rm dec}$.  For sufficiently small kick velocities ($V_{\rm k}\lesssim$ 10 km s$^{-1}$), equation~\ref{eq:rbow}  no longer provides a  good description for $R_{\rm s}$ and a spherically symmetric calculation becomes more appropriate \citep{2013ApJ...768..113M,2013MNRAS.431.2737M}.   Such cavities can be as large as fractions of a parsec or more, giving rise to shock deceleration months after the sGRB, and resulting in a much lower afterglow luminosity that could easily go undetected. However, given that NS/NS and NS/BH binaries are typically bestowed velocities in excess of $100$ km s$^{-1}$ \citep[e.g.][]{1998ApJ...499..520F}, we do not expect that this will commonly occur. Finally, we note that $R_{\rm s}$ represents a {\it lower limit} to the bow-shock distance scale, which can be enhanced by factors of two or more depending on the orientation of the binary within the cavity.  

\begin{figure}[t]
\centering\includegraphics[width=0.8\linewidth,clip=true]{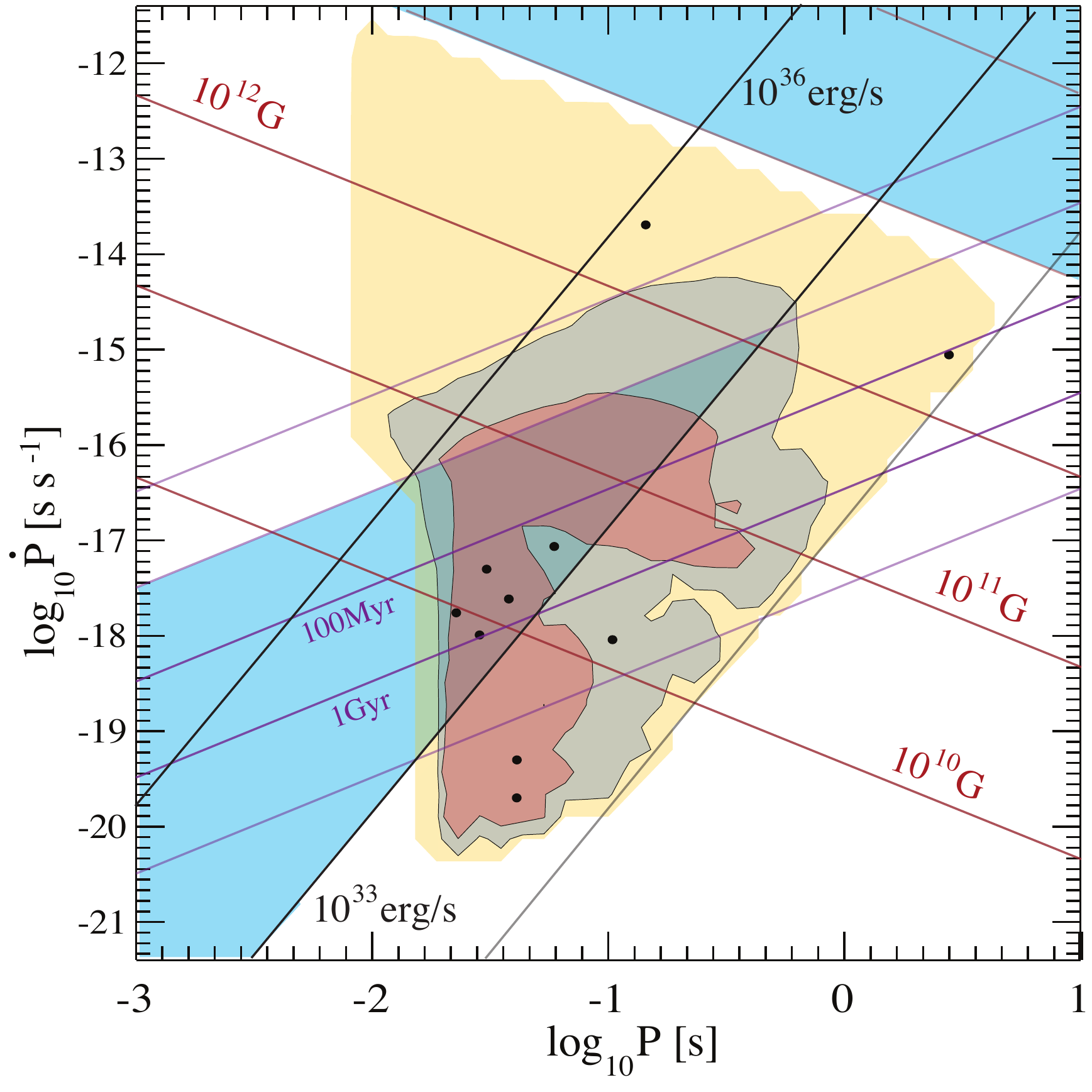}
\caption{$P$-$\dot{P}$ diagram with pulsar lifetimes $\tau_{\rm p}$ ({\it purple} lines), magnetic fields $B$ ({\it red} lines), and spin-down luminosities $L_{\rm p}$ ({\it black} lines). The {\it light blue} shaded regions indicate where the condition $R_{\rm s} > R_{\rm dec}$ is satisfied, with the additional constraint that $\tau_{\rm p} > 10$ Myr for pulsars with fields smaller than $\sim 10^{13}$ G. The remaining contours are reproduced from the {\it HP} NS/NS population synthesis model of \cite{2011MNRAS.413..461O}, and represent 1- ({\it red}), 2- ({\it purple}), and 3- ({\it yellow}) $\sigma$ probability for a NS/NS to host a pulsar companion within the given contour. Black dots represent observed pulsars in binaries (10 pulsars in 9 binaries; see \cite{2011MNRAS.413..461O} for details). Depending on the details of population synthesis, up to $\approx 50\%$ of NS/NS binaries will satisfy $R_{\rm s} > R_{\rm dec}$.}
\label{fig:ppdot}
\end{figure}

In this {\it Letter} we have shown that, under certain conditions, the behavior of sGRB afterglows can be altered from the predictions of the standard theory, independently of long-lasting jet activity. In Fig. \ref{fig:ppdot} we translate these conditions to a $P$-$\dot{P}$ diagram, where $P$ and $\dot{P}$ are the pulsar period and its time derivative, respectively. As discussed in previous sections, the region of interest separates into two branches: low field pulsars with long spin-down time scales and high field pulsars ($B\gtrsim10^{13}$ G). Low field pulsars depend on their spin to generate  the luminosity required to ensure $R_{\rm s}>R_{\rm dec}$,  so that  an additional constraint should be  that $\tau_{\rm p} \gtrsim \tau_{\rm GW}$. In Fig.  \ref{fig:ppdot}, we have truncated the region of interest to exclude low field pulsars with $\tau_{\rm p} < 10$ Myr. To get a sense for the likelihood that such conditions can actually be realized, we have included population contours from the NS/NS population synthesis modelling of \cite{2011MNRAS.413..461O}. With the caveat that the uncertainties of population synthesis are significant, a sizeable fraction of NS/NS binaries lie within the region of interest for the most likely model of \cite{2011MNRAS.413..461O}, and all of these systems are in the low field, long lived pulsar branch.

As we have demonstrated, the spindown luminosity always dominates the growth of the standoff shock radius provided that  $B\lesssim 10^{13}$ G. If this condition is  not satisfied, then the pulsar lifetime $\tau_{\rm p}$ becomes irrelevant because  most  of the energy injection into the cavity takes places as the  binary merges. However, the preceding argument is only valid under the assumption that no significant magnetic field decay occurs. The processes of pulsar field decay are controversial, with suggested time scales between 100 kyr and 1 Gyr \citep{2011MNRAS.413..461O,2012NewA...17..594C,2014AN....335..262I,2014MNRAS.437.3863K}. Given these uncertainties,  we urge caution in the application of these results for high field pulsars.

\begin{figure}[]
\centering\includegraphics[width=0.28\linewidth,clip=true]{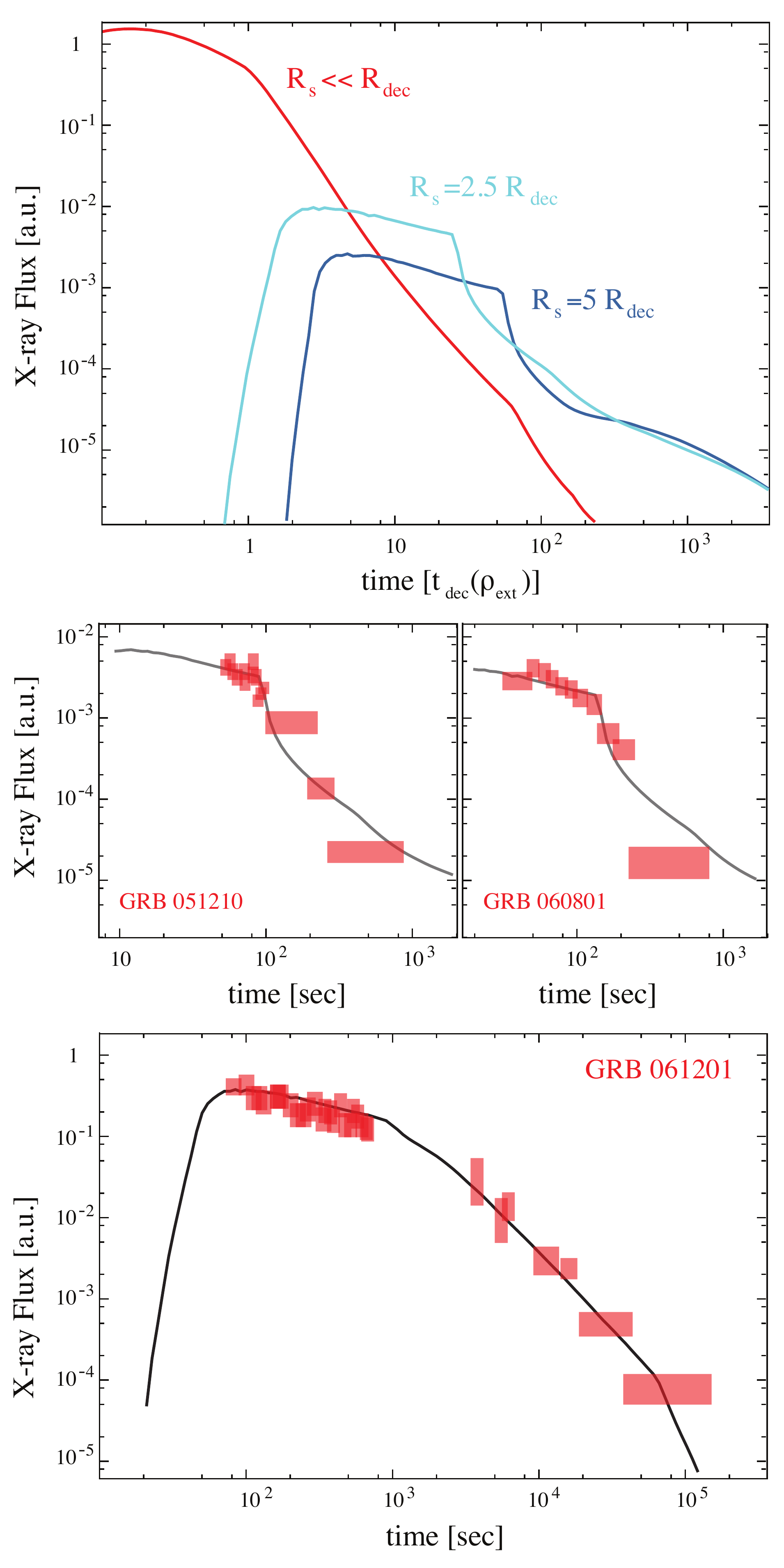}
\caption{X-ray afterglows from sGRBs taking place inside cavities. {\it Top Panel}: Three examples are depicted where the sGRB blastwave properties are unchanged but the external media are altered. For the {\it red} curve, the X-ray afterglow is calculated for a sGRB blastwave expanding into a constant density medium with $\rho_{\rm ext}$. The typical deceleration radius is $R_{\rm dec}(\rho_{\rm ext})$, which corresponds to an observed time $t_{\rm dec}$.  The effects of  a pulsar wind cavity on the sGRB afterglow are clearly demonstrated by comparing  the {\it red} and  {\it blue} curves. In the latter cases, the blastwave expands unhindered within the cavity  until it reaches the thin shell  placed at $R_{\rm s}=2.5R_{\rm dec}(\rho_{\rm ext})$ ({\it light blue}) and $R_{\rm s}=5R_{\rm dec}(\rho_{\rm ext})$ ({\it dark blue}), respectively.  {\it Middle panels:} A blastwave interacting  with a   pulsar wind cavity with $R_{\rm s}=2.5R_{\rm dec}$ provides  a reasonable explanation for  the X-ray ligthcurves of GRB 051210 ($t_{\rm dec, cav}=5$s) and GRB 060801 ($t_{\rm dec, cav}=8$s). {\it Bottom panel:} GRB 061201 can be modeled with $R_{\rm s} \approx R_{\rm dec}$ and $t_{\rm dec,cav}=26$s. These one dimensional, hydrodynamic models take on the density profile of the pulsar wind cavity  in the polar direction and provide a good description of the observed afterglow lightcurve as long as the flow is confined to a jet with opening angle $\theta_{\rm jet} \lesssim 20^{\circ}$. In all calculations we assume  $\Gamma_0=10$, $\Delta R/R_{\rm s} = 0.1$, and that  $\epsilon_e$ and $\epsilon_B$  remain constant.}
\label{fig:aft}
\end{figure}

It is clear from the analysis presented here that the afterglow lightcurve can be strongly modified by the presence of a pulsar within the system.  This implies that we cannot be too specific about the exact times at which we expect to see transitions in the afterglow, but if and when we do see these transitions, they can be fairly constraining on the properties of the system. Our results indicate, however,  that  for canonical values the expected range of time delays are consistent with  {\it Swift}  observations of sGRB afterglows (Fig.~\ref{fig:aft}), particularly when accounting for the instrument dead-time \citep[e.g.][]{2013MNRAS.430.1061R}. Although more elaborate modelling may be needed to explain the few extreme fading X-ray afterglows, this simplified scenario ameliorates the need for a long-lasting central engine, for which it is hard to see how it might naturally generate the prompt emission \citep{2014arXiv1404.0383M}.

\acknowledgements
We thank C. Dermer, W. Lee,  A. Van Der Horst, and O. Pejcha for discussions and acknowledge financial support from the Packard Foundation, NSF (AST0847563), PAPIIT-UNAM (IA101413-2), and UCMEXUS (CN-12-578).


\end{document}